\documentclass[final]{agujournal2019}
\usepackage{url} 
\usepackage[inline]{trackchanges} 
\usepackage{soul}
\usepackage{graphicx}
\usepackage{booktabs,amsmath}
\usepackage{graphicx}
\usepackage{subcaption}

\newif\ifboldrevisions
\boldrevisionsfalse 

\newcommand{\rev}[1]{\ifboldrevisions\textbf{#1}\else#1\fi}
\newcommand{\ec}[1]{{\color{blue}#1}}
\newcommand{\ai}[1]{{\color{black}#1}}
\newcommand{\tb}[1]{{\color{blue}#1}}

\journalname{Space Weather}

\begin{document}

%
%


\title{Verification of the NOAA Space Weather Prediction Center solar flare forecast (1998-2024)}

%
%




\authors{Enrico Camporeale\affil{1,2}, Thomas E. Berger\affil{3}}

\affiliation{1}{School of Physical and Chemical Sciences, Queen Mary University of London, UK}
\affiliation{2}{Space Weather Technology, Research and Education Center, University of Colorado, Boulder, CO, USA}
\affiliation{3}{National Center for Atmospheric Research, High Altitude Observatory, Boulder, CO, USA}





\correspondingauthor{Enrico Camporeale}{enrico.camporeale@colorado.edu}



\begin{keypoints}

 \item SWPC solar flare forecasts do not outperform zero-cost baselines such as persistence and climatology over a 26-year validation period.
  \item SWPC solar flare forecasts are poorly calibrated and produce excessive false alarms, particularly for X-class flares and in “all-clear” scenarios.
  \item We recommend the operational adoption of modern machine learning methods and routine verification of forecast skill.
\end{keypoints}

%
%

%
%


\begin{abstract}
\ai{
The NOAA Space Weather Prediction Center (SWPC) issues the official U.S. government forecast for M-class and X-class solar flares, yet the skill of these forecasts has never been comprehensively verified. In this study, we evaluate the SWPC probabilistic flare forecasts over a 26-year period (1998–2024), comparing them to several zero-cost and statistical baselines including persistence, climatology, Naive Bayes, and logistic regression. We find that the SWPC model does not outperform these baselines across key classification and probabilistic metrics and exhibits severe calibration issues and high false alarm rates, especially in high-stakes scenarios such as detecting the first flare after extended quiet periods.}\rev{\tb{These findings demonstrate the need for more accurate and reliable eruption forecasting models which we suggest should be based on modern data-driven methods. The findings also provide a standard against which any proposed eruption prediction system should be compared. We suggest that space weather forecasters regularly update and publish analyses like the one demonstrated here to provide up-to-date standards of accuracy and reliability against which to compare new methods.}}
\end{abstract}

\section*{Plain Language Summary}
\ai{The U.S. government regularly provides space weather forecasts, including predictions of X-rays from solar eruptions—sudden bursts of energy from the Sun that can harm satellites, power grids, and astronauts. Many people and organizations use these official forecasts when planning activities that depend on space weather. In this study, we looked closely at how accurate those forecasts have been over the last 26 years. We compared them to very simple methods, like just assuming tomorrow will be similar to today, or using past averages. Surprisingly, we found that the official forecasts were no better than these basic methods. In some important situations—like when the Sun has been quiet for weeks and a powerful flare might occur—the forecasts often failed. This is a serious concern, especially for astronaut safety. We suggest that modern data-driven techniques, like machine learning, could produce better predictions than current techniques and should be prioritized for development. We also encourage space weather forecasters to regularly test and publish how well their predictions are working.}

%
%
\onecolumn

%


%
%
%
%
\ai{}
\section{Introduction}
\ai{
Solar eruptions are among the most energetic phenomena in the heliosphere, capable of releasing up to $10^{32}$ erg of energy in the form of electromagnetic radiation, energetic particles, and plasma ejections \cite{benz2017flare}. These events are key drivers of space weather, with the potential to disrupt satellite operations, degrade radio communications, and endanger astronauts in space} \ec{ \cite{schwenn2006space, schrijver2015understanding, manuela2021space}. }
\ai{Accurate forecasting of solar eruption occurrence is therefore of critical importance for operational space weather services.}

\ec{The first indication of a solar eruption to reach Earth is the \rev{electromagnetic} radiation generated primarily in X-ray and Extreme Ultraviolet (EUV) photons. Colloquially referred to as ``solar flares'', solar eruption photonic emission is classified according to its peak soft X-ray flux in the 1--8 \AA{} band as observed by the NOAA/GOES satellites. The original flare classification system included Common (C), Medium (M), and eXtreme (X) classes. Later classifications included lower-level X-ray emission in the A and B classes. Each class represents a tenfold increase in peak 1--8 \AA\ flux. Of particular concern for space weather are M-class and X-class flares, which can cause moderate to severe high frequency (3--30 MHz) radio and radar interference on the sunlit side of the Earth and are often associated with large plasma eruptions (so-called ``Coronal Mass Ejections'' or CMEs) that, if they are Earth-directed, can cause moderate to extreme geomagnetic storms \cite{hill2005noaa, gopalswamy2018extreme}.}

\ai{Several solar flare catalogs are maintained to support research and operational activities. These include the NOAA event lists derived from GOES X-ray data, the Heliophysics Event Knowledgebase (HEK)} \ec{ \cite{hurlburt2012heliophysics}}\ai{, and curated flare event databases such as those produced by the Solar Data Analysis Center (SDAC) and the Kanzelhöhe Solar Observatory} \ec{ \cite{potzi2015real, chen2024solar}. A solar flare catalog based on NASA Solar Dynamics Observatory (SDO) Atmospheric Imaging Assembly (AIA) EUV images was recently presented in \citeA{van2022solar}.} \ai{These catalogs are foundational resources not only for retrospective event studies but also for the development and validation of empirical and machine learning-based flare forecasting models} \ec{ \cite{leka2018solar, leka2019comparison, florios2018forecasting, camporeale2019challenge,georgoulis2024prediction}.} \ec{Recently, \citeA{berretti2025asr} produced a new catalog called ASR (Archival Solar Flares), which includes a method to locate flare locations on the Sun and link them to photospheric active regions using SDO data. In this work, we used the ASR v1.0.0 (released on March 12, 2025) available on \url{https://github.com/helio-unitov/ASR_cat/releases/download/v1.1/f_1995_2024.csv}. The dataset covers flares that occurred during the period 2002-2024.}

\ec{
The official US government source of solar eruption/flare forecasts is the NOAA Space Weather Prediction Center (SWPC) in Boulder, Colorado, which provides operational (i.e., continuously available with robust backup capabilities) forecasts for M- and X-class flares. 
Three-day solar flare forecasts are issued every 12 hours and summarized daily (the latest 3-day forecast is available at \url{https://www.swpc.noaa.gov/products/3-day-forecast} and summaries are available at \url{https://www.swpc.noaa.gov/products/report-and-forecast-solar-and-geophysical-activity}). The 3-day flare forecasts use the SWPC-specific ``Radio Blackout'' or R-scale for flare classification where an R1 flare corresponds to an M1--4.9 X-ray peak, R2 corresponds to an M5--9.9 flare, and R3--R5 classes correspond to X-class flares. In contrast, the summary report and forecast uses the M- and X-class nomenclature. In spite of the differing nomenclature both forecast products consist of three integer numbers that are the forecast probability of the occurrence of one or more M- or X-class flares for each of the next 3 days (see \url{https://www.swpc.noaa.gov/noaa-scales-explanation} for an explanation of NOAA/SWPC space weather scales),. The SWPC forecast is used by a range of stakeholders, including satellite operators, airlines, air traffic controllers, radar operators, and defense agencies.} \ai{Despite their central role in space weather forecasting, the methodology behind SWPC's flare probability forecasts is not publicly documented in detail, and, to the best of the authors' knowledge, no comprehensive verification study of their accuracy and reliability, or skill, has been published} \ec{ since the work of \citeA{crown2012validation}, which covered the period of 1996-2008. More recent studies, such as \citeA{leka2019comparison}, verified the forecasts over only a short period of time (2016-2017).}

\ec{SWPC \rev{flare} forecast methodology is not described on the NOAA website, however, according to \citeA{leka2019comparison}} \ai{the methodology begins with classifying active regions using the McIntosh scheme \cite{mcintosh1990classification} and assigning flare probabilities based on historical flaring rates established for each McIntosh class (i.e., a climatological baseline); these probabilities are then modified by human forecasters based on region evolution, recent flaring activity, and expert judgment, and the resulting region-specific probabilities are aggregated into a full-disk forecast, which may be further adjusted considering the activity of recently rotated-off or returning active regions and, when available, supplemented by additional model outputs; the initial forecast is issued at 22:00 UTC for the next day and incorporated into the official three-day forecast released at 00:30 UTC and updated at 12:30 UTC.}

\ai{The goal of this study is to systematically verify the skill of SWPC’s probabilistic forecasts for M-class and X-class flares} \ec{ over the period \rev{1998-2024 (26 years of data)}.} \ai{We compare forecast probabilities with flare occurrence records from NOAA’s GOES-based catalog, employing a number of standard metrics for classification. This analysis contributes both to operational forecasting assessment and to the broader objective of benchmarking baseline models for flare prediction, which is increasingly important in the era of data-driven and machine learning-based space weather modeling.}

\section{Data}

\ai{This study uses three primary data sources: the official daily probabilistic forecasts issued by NOAA's Space Weather Prediction Center (SWPC), the flare occurrence records from the NOAA event reports, and the ASR catalogue. 

NOAA/SWPC issues daily forecasts for the probability of at least one M-class or X-class flare occurring within the next 1, 2, and 3 days. Historical records of these forecasts, dating back to 1996, are available from the SWPC data archive at~\url{ftp://ftp.swpc.noaa.gov/pub/warehouse/}. These records are stored as annual text files. We parsed the forecasts from each file and compiled them into a unified CSV format for further analysis. Each row in the resulting dataset corresponds to a unique date and contains the forecast probabilities for M-class and X-class flares issued on that day \rev{(six numbers: 1, 2, 3-days ahead for M- and X-class).}
}
\ai{
Flare occurrence data were obtained from two sources. For the years 2002--2024, we used the ASR flare catalogue, which includes flare events curated from the GOES X-ray flux records and incorporates the official NOAA event data (\url{https://github.com/helio-unitov/ASR_cat}). For the years 1996--2001, flare records were retrieved directly from the NOAA SWPC event reports, available at~\url{ftp://ftp.swpc.noaa.gov/pub/indices/events/}. 

To perform the validation, we constructed a binary "ground truth" label for each day in the dataset. A day is labeled as positive (1) for a given class (M or X) if at least one flare of that class occurred during that day (UTC). Otherwise, it is labeled as negative (0). This daily binarization of the flare records aligns with the forecast product, which specifies probabilities of at least one event per day.}

\ai{The merged dataset spans 10,338 unique days.} \ec{As explained in Section \ref{sec:training}, we do not perform a statistical analysis of the first 17 months (1996-08 to 1997-12) and instead use those data as a training buffer for baseline models. For the remaining period (1998-2024),} \ai{Table~\ref{tab:flare_counts} summarizes the total number of positive and negative instances for both M-class and X-class flares} \ec{ (once again, those are the number of days with at least one flare, the information about how many flares occur in a day is not relevant).} \ai{ The distribution is highly imbalanced, particularly for X-class flares, which occur far less frequently than M-class flares.}

\begin{table}[ht]
\centering
\caption{Total counts of positive and negative days for M-class and X-class flares for the period 1998--2024 (9,828 days).}
\label{tab:flare_counts}
\begin{tabular}{lcc}
\toprule
 & \textbf{M-class} & \textbf{X-class} \\
\midrule
Positive days (label = 1) & 2021 & 254 \\
Negative days (label = 0) & 7807 & 9574 \\
Class imbalance ratio (neg/pos) & 3.86 & 37.69 \\
\bottomrule
\end{tabular}
\end{table}

\begin{figure}[ht]
\centering
\includegraphics[width=0.9\textwidth]{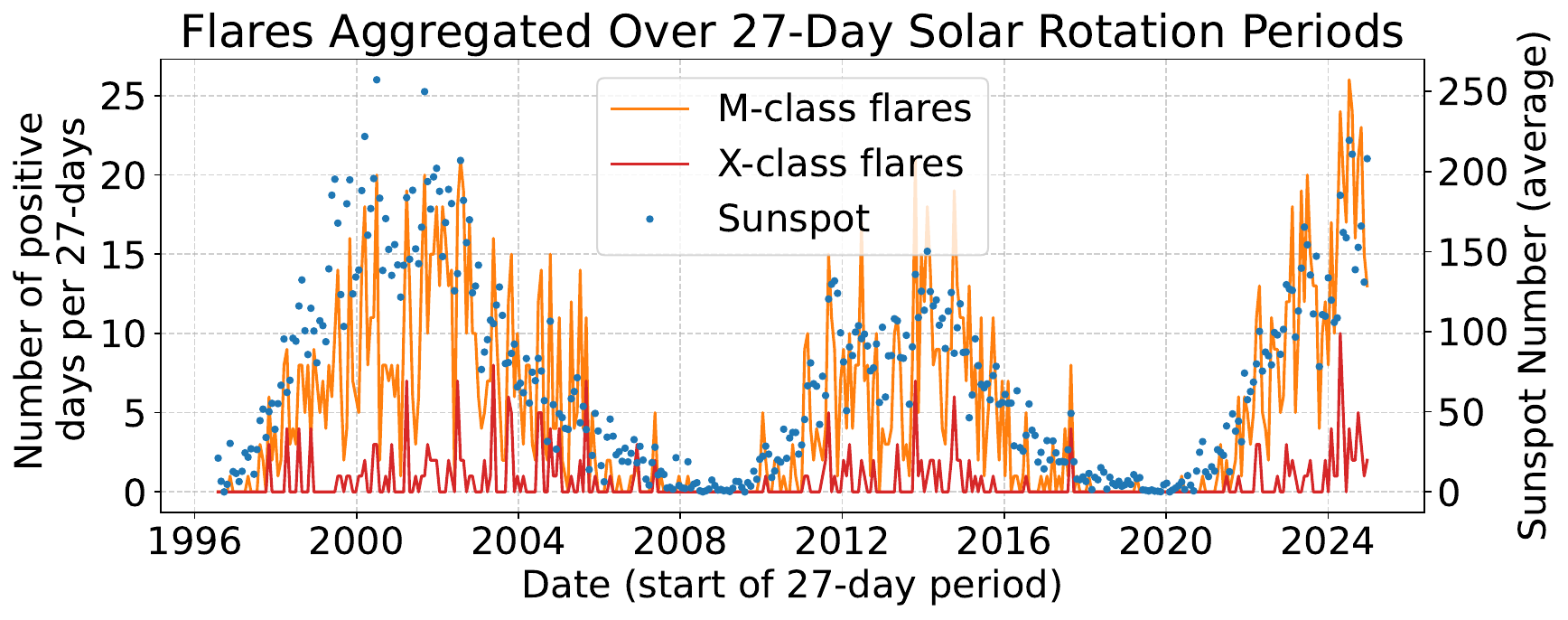}
\caption{Number of positive days (M-class or X-class) in a 27-day sliding window, plotted over the study period. Blue dots indicate the 27-days average of sunspot numbers. Modulation due to the approximately 11-year solar magnetic activity cycle is clearly visible.}

\label{fig:cycle_plot}
\end{figure}

\ai{To illustrate long-term solar cycle variability, we computed a 27-day rolling} \ec{ (non-overlapping)} \ai{sum of positive flare days and plotted it as a function of time. Figure~\ref{fig:cycle_plot} shows the number of days within each 27-day window during which at least one M-class or X-class flare occurred. This smoothed activity index clearly tracks the solar cycle, with pronounced peaks near solar maxima and extended periods of low activity near solar minima} \ec{ \rev{(27-days average sunspot number shown as blue dots)}.} \ai{The 27-day window was chosen to approximate the solar rotation period, capturing active-region recurrence patterns.}

\begin{figure}[ht]
\centering
\includegraphics[width=0.8\textwidth]{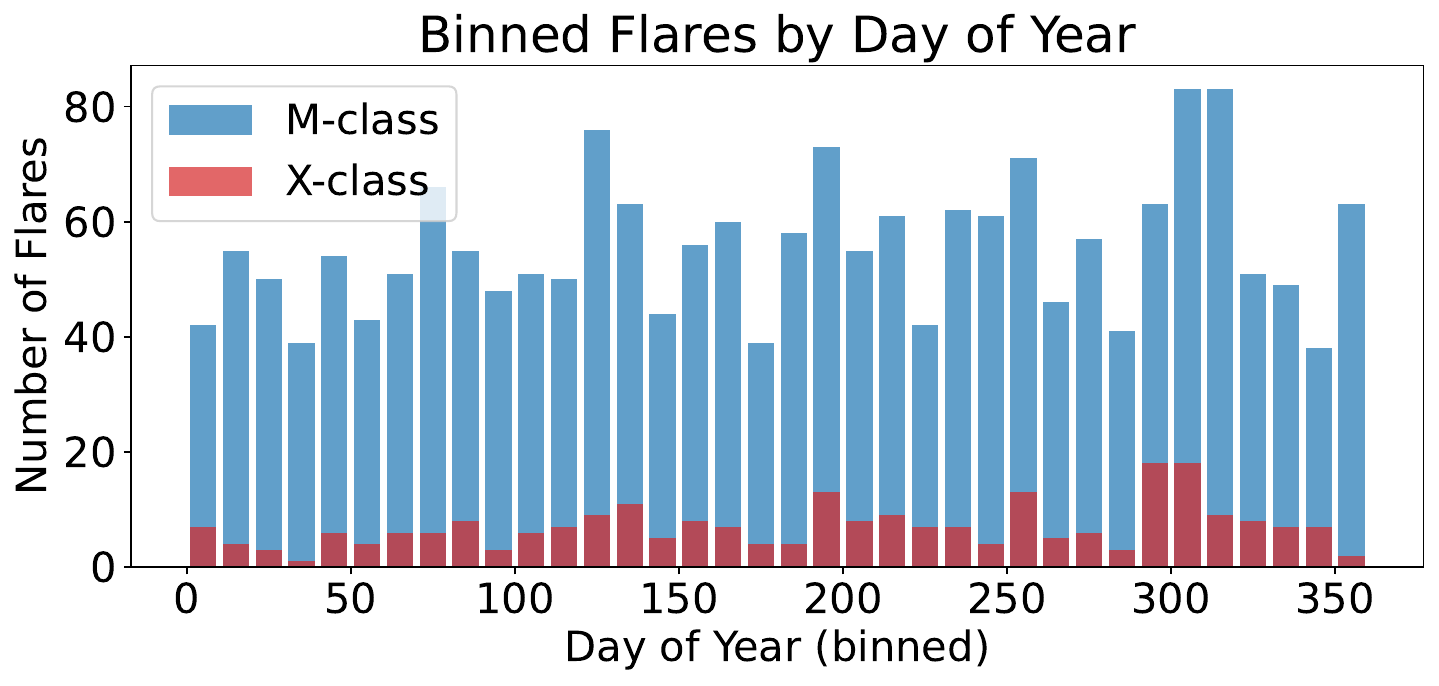}
\caption{Total number of M-class and X-class flares as a function of day of year, aggregated over the entire dataset.}
\label{fig:flares_vs_doy}
\end{figure}

\ai{We also analyzed the seasonal distribution of flares. Figure~\ref{fig:flares_vs_doy} shows the number of M-class and X-class flares as a function of the day of the year, aggregated over all years. This seasonal profile highlights the variability of flare activity, which is modulated by the solar cycle but does not show strong annual periodicity.}

\begin{figure}[ht]
\centering
\includegraphics[width=0.8\textwidth]{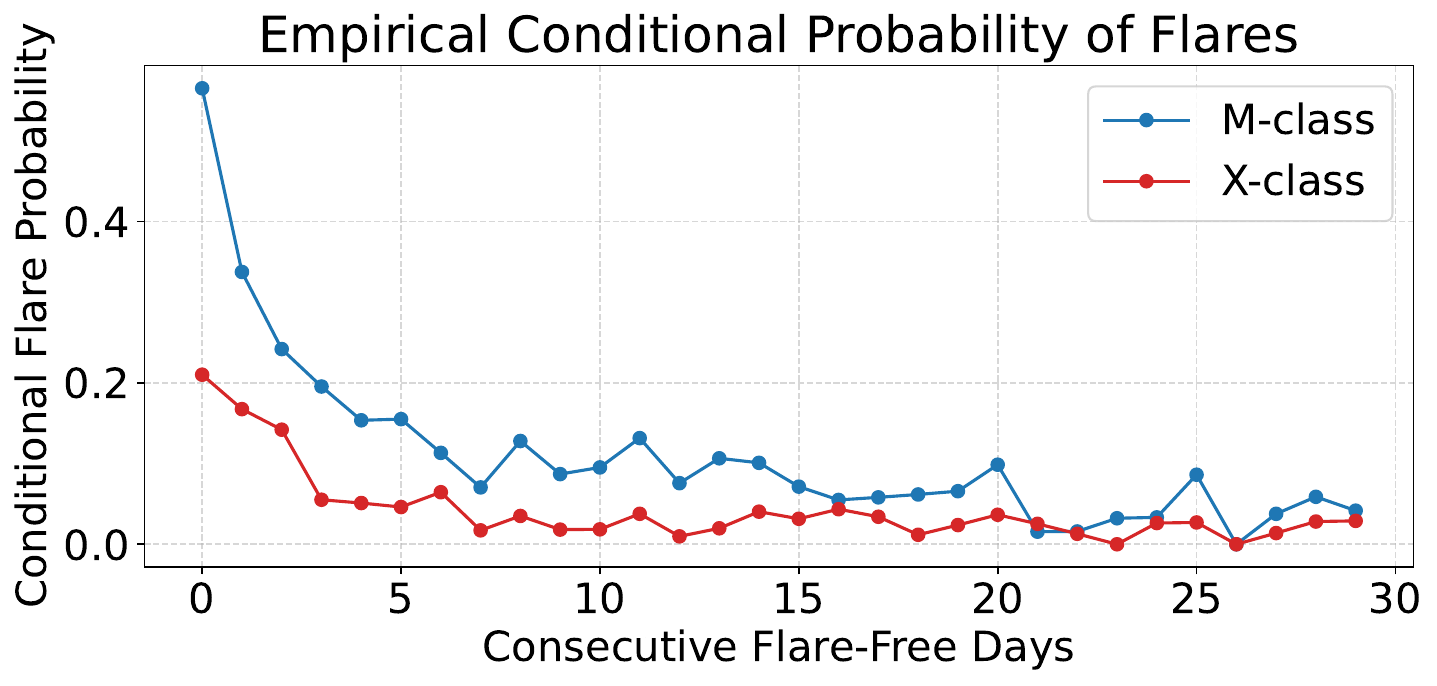}
\caption{Conditional probability of observing an M-class or X-class flare given $n$ consecutive prior days without such a flare.}
\label{fig:conditional_prob}
\end{figure}

\ai{In addition, we computed the conditional probability that a flare would occur as a function of the number of preceding flare-free days. This quantity is useful to assess the memory effects or persistence in flare activity. Figure~\ref{fig:conditional_prob} displays the empirical conditional probabilities for both M-class and X-class flares.

For M-class flares, the conditional probability of occurrence is greater than 50\% if a flare occurred on the previous day} \ec{ (zero prior flare-free days),} \ai{ indicating strong short-term persistence in flare activity. In contrast, the conditional probability for X-class flares is around 20\% after a flare day. For both flare classes, the conditional probability decays with increasing flare-free duration and appears to plateau after approximately 10 days, suggesting limited memory beyond that scale.}

\section{Methodology}

\ai{To evaluate the skill of the SWPC probabilistic forecasts for M-class and X-class flares, we adopt a suite of metrics applicable to both deterministic and probabilistic forecasts. The SWPC predictions are issued as probabilities, so we convert them into binary (yes/no) predictions using a probability threshold and compute standard classification scores. In addition, we assess the forecasts directly in their probabilistic form using proper scoring rules and reliability diagrams.}

\subsection{Confusion Matrix and Binary Metrics}

\ai{Given a threshold $\theta$, we define a flare-day prediction if the forecast probability $p \geq \theta$. This thresholding converts the probabilistic forecasts into binary predictions, which can be compared to the ground-truth binary labels (flare or no flare) for each day.

From this comparison, we construct a confusion matrix:

\begin{itemize}
    \item \textbf{True Positive (TP)}: the forecast predicted a flare and a flare occurred.
    \item \textbf{False Positive (FP)}: the forecast predicted a flare but no flare occurred.
    \item \textbf{True Negative (TN)}: the forecast predicted no flare and no flare occurred.
    \item \textbf{False Negative (FN)}: the forecast predicted no flare, but a flare occurred.
\end{itemize}

Based on these quantities, we compute the following deterministic performance metrics:}

\begin{itemize}
    \ai{
    \item \textbf{Accuracy (ACC)}: the proportion of correct predictions (both flare and no-flare days):
    \[
    \mathrm{ACC} = \frac{\mathrm{TP} + \mathrm{TN}}{\mathrm{TP} + \mathrm{FP} + \mathrm{TN} + \mathrm{FN}}
    \]

    \item \textbf{Precision (PREC)}: the proportion of predicted flare days that were correct} \ec{ (that is, the likelihood that a flare occurs, when the model predicts a flare):}
    \ai{
    \[
    \mathrm{PREC} = \frac{\mathrm{TP}}{\mathrm{TP} + \mathrm{FP}}
    \]

    \item \textbf{Recall (REC)} or \textbf{Probability of Detection (POD)}: the proportion of actual flare days that were correctly predicted:
    \[
    \mathrm{REC} = \mathrm{POD} = \frac{\mathrm{TP}}{\mathrm{TP} + \mathrm{FN}} = \frac{\mathrm{TP}}{\mathrm{P}}
    \]

    \item \textbf{False Alarm Ratio (FAR)}: the fraction of predicted flare days that were false alarms} \ec{ (that
is, the likelihood that a flare does NOT occur, when the model predicts a flare):
    \[
    \mathrm{FAR} = 1-PREC = \frac{\mathrm{FP}}{\mathrm{TP} + \mathrm{FP}}
    \]
}
    \item \textbf{False Positive Rate (FPR)}: the fraction of positive predictions that are incorrect relative to the total number of negative events:
    \[
    \mathrm{FPR} = \frac{\mathrm{FP}}{\mathrm{FP + TN}}
    \]
    
\ai{
    \item \textbf{F1-score (F1)}: the harmonic mean of precision and recall, which balances false positives and false negatives:
    \[
    \mathrm{F1} = 2 \cdot \frac{\mathrm{PREC} \cdot \mathrm{REC}}{\mathrm{PREC} + \mathrm{REC}} = \frac{2 \cdot \mathrm{TP}}{2 \cdot \mathrm{TP} + \mathrm{FP} + \mathrm{FN}}
    \]

    \item \textbf{Critical Success Index (CSI)}: the proportion of observed and/or predicted events that were correctly predicted:
    \[
    \mathrm{CSI} = \frac{\mathrm{TP}}{\mathrm{TP} + \mathrm{FP} + \mathrm{FN}}
    \]
}\ec{
    \item \textbf{True Skill Statistic (TSS)}: the difference between the probability of detection and the false positive rate:
    \[
    \mathrm{TSS} = \frac{\mathrm{TP}}{\mathrm{TP} + \mathrm{FN}} - \frac{\mathrm{FP}}{\mathrm{FP} + \mathrm{TN}}
    \]
}
\ai{
    \item \textbf{Heidke Skill Score (HSS)}: the fractional improvement of the forecast over random chance:
    \[
    \mathrm{HSS} = \frac{2 (\mathrm{TP} \cdot \mathrm{TN} - \mathrm{FN} \cdot \mathrm{FP})}
    {(\mathrm{TP} + \mathrm{FN})(\mathrm{FN} + \mathrm{TN}) + (\mathrm{TP} + \mathrm{FP})(\mathrm{FP} + \mathrm{TN})}
    \]
}    
\end{itemize}

\subsection{ROC Curve and Optimal Threshold}

\ai{To evaluate the performance of the probabilistic forecasts across possible thresholds, we compute the Receiver Operating Characteristic (ROC) curve. The ROC curve plots the True Positive Rate (POD) against the False Positive Rate (FPR) as the decision threshold varies. The area under the ROC curve (AUC) quantifies the ability of the forecast to discriminate between flare and no-flare days; a perfect forecast has AUC = 1, while a no-skill forecast has AUC = 0.5.} \ec{ Note that the TSS is the \rev{maximum} vertical distance between the ROC curve and the diagonal line.}




\subsection{Probabilistic Forecast Evaluation}
\ai{
In addition to evaluating binary classifications, we assess the quality of the raw probability forecasts. Two standard metrics are used:}

\begin{itemize}
\ai{
    \item \textbf{Brier Score (BS)}: the mean squared error between forecast probabilities $p_i$ and binary outcomes $y_i \in \{0,1\}$:
    \[
    \mathrm{BS} = \frac{1}{N} \sum_{i=1}^N (p_i - y_i)^2
    \]
    Lower Brier Scores indicate more accurate and better-calibrated probabilistic forecasts.}

\ai{    \item \textbf{Reliability Diagram}: this plot compares the forecast probability to the observed frequency of flare occurrence. A well-calibrated forecast will fall along the diagonal; systematic deviations indicate overconfidence (below the diagonal) or underconfidence (above it).}
\end{itemize}

\subsection{Baseline Models and Context for Evaluation Metrics}
\ai{
To contextualize the performance of the SWPC forecasts, we compare them to several simple, interpretable models that require minimal computational cost and leverage only readily available solar indicators. These \textit{zero-cost baselines} serve as reference points to determine whether operational forecasts provide added value beyond naive or climatological heuristics.}

 \subsubsection{Persistence Model.} \ai{The persistence model uses flare occurrence on the current day to predict flare occurrence on subsequent days. For each day $D$, the observed flare activity (M- or X-class) is used as the prediction for day $D+1$, $D+2$, and $D+3$. While conceptually simple, this model can yield non-trivial skill during active solar periods when flare events are temporally clustered.
}

\subsubsection{Climatology-Based and Statistical Models.} \ai{We define three additional baselines that all rely on the same two input features: }
\begin{itemize}
\ai{
    \item $x_1$: number of consecutive flare-free days prior to the forecast day;
    \item $x_2$: sunspot number on the forecast day.}
\end{itemize}
 \ai{
These features capture aspects of solar activity relevant to flare occurrence and are available in real time, making them suitable for operational use.}
\ai{
The three models differ in how these inputs are used:}
\begin{itemize}
\ai{
    \item \textbf{Empirical Climatology:} For each combination of flare-free days and sunspot number $(x_1, x_2)$, we compute the empirical probability of a flare based on historical records. This conditional probability serves directly as the predicted probability of flare occurrence.
 The features are discretized into bins: $(0,1,2,\ldots,20,>20)$ for $x_1$, and $(0,10,20,\ldots,200,>200)$ for $x_2$. For each bin pair $(b_1, b_2)$, we compute:

\begin{equation}
    P_\text{clim}(y=1 \mid x_1 \in b_1, x_2 \in b_2) = \frac{N_{\text{flare}}(b_1, b_2)}{N_{\text{total}}(b_1, b_2)},
\end{equation}

where $N_{\text{flare}}(b_1, b_2)$ is the number of days with a flare in the bin pair, and $N_{\text{total}}(b_1, b_2)$ is the total number of days in that bin. This model reflects the empirically observed frequency of flares conditioned on recent activity and sunspot levels.
    
   \item \textbf{Naive Bayes Classifier (NB):} 
Naive Bayes is a probabilistic classifier that assumes conditional independence between the input features given the class label $y \in \{0, 1\}$:

\begin{equation}
    P(y=1 \mid x_1, x_2) = \frac{P(y=1) P(x_1 \mid y=1) P(x_2 \mid y=1)}{\sum_{y' \in \{0, 1\}} P(y') P(x_1 \mid y') P(x_2 \mid y')}.
\end{equation}

The class priors $P(y=1)$ and $P(y=0)$ are computed from the dataset. The feature likelihoods $P(x_1 \mid y)$ and $P(x_2 \mid y)$ can be estimated either using histograms or kernel density estimates.} \ec{Despite the strong assumption of feature independence (which is obviously not valid, given that sunspot number is not independent from number of consecutive flare-free days), Naive Bayes performs reasonably well. In this work, we use the Multinomial Naive Bayes implementation of scikit-learn \cite{pedregosa2011scikit}.}
    \ai{
    \item \textbf{Logistic Regression (LR):} A linear probabilistic model in which the log-odds of flare occurrence is modeled as a function of the two input features. It provides a parametric alternative to the binned empirical climatology and can generalize better in regions of low data density.

\begin{equation}
    \log \left( \frac{P(y=1 \mid x_1, x_2)}{1 - P(y=1 \mid x_1, x_2)} \right) = \beta_0 + \beta_1 x_1 + \beta_2 x_2.
\end{equation}

Equivalently, the probability is given by the sigmoid function:

\begin{equation}
    P_\text{LR}(y=1 \mid x_1, x_2) = \frac{1}{1 + \exp(-(\beta_0 + \beta_1 x_1 + \beta_2 x_2))}.
\end{equation}

The model parameters $\beta_0, \beta_1, \beta_2$ are learned by maximizing the likelihood of the observed training data. Logistic regression captures the interaction between features in a parametric form and can generalize to unseen combinations better than the binned climatology model.}

\ai{ \item \textbf{Baseline Average:}
\rev{In addition to evaluating each model individually, we also compute a \textit{baseline average} forecast, defined as the arithmetic mean of the predicted probabilities from four models: climatology, Naive Bayes, logistic regression, and persistence. 
Averaging forecasts is a common and effective ensemble strategy that helps reduce overfitting and model-specific biases. By averaging across these methods, the ensemble benefits from the diversity of its members and often achieves better calibration, lower variance, and improved robustness compared to any single model. This is particularly valuable in rare-event forecasting, where individual models may exhibit erratic behavior due to limited data.}
}
\end{itemize}

\subsection{Training Strategy}\label{sec:training}
\ai{
\rev{To ensure a fair and realistic comparison across all forecasting methods, we adopt a training strategy that strictly avoids the use of future information when generating predictions. Specifically, for the statistical and machine learning models (climatology, logistic regression, and Naive Bayes), we only use training data that would have been available up to the time each forecast was made. This simulates an operational scenario and prevents any leakage of future information into the training set.}

\rev{We reserve the first 17 months of data (from 1996-08 to 1997-12) as a buffer period, used solely for model initialization and excluded from the evaluation. Beginning in January 1998, all models are retrained monthly: that is, each month’s forecasts are produced using a model trained only on data from all previous months. This rolling retraining approach reflects how an operational system might continuously incorporate new data while remaining causally consistent.}

\rev{While this retraining scheme is particularly important for machine learning models such as logistic regression and Naive Bayes to avoid overfitting, it is, in principle, less critical for the climatology model. Climatology is typically defined as a static method based on long-term averages. However, for methodological consistency, we apply the same monthly retraining procedure to the climatology model as well.}
\rev{Persistence, on the contrary, does not require any training.
}
}
\subsection{Thresholding for Binary Metrics.}
\ec{The SWPC forecast and all of the baseline models (except for persistence)} \ai{ produce probabilistic outputs, i.e., $P(y=1 \mid x_1, x_2)$. To evaluate classification performance using binary metrics, we must apply a decision threshold $\theta$:

\begin{equation}
    \hat{y} = 
    \begin{cases}
        1 & \text{if } P(y=1 \mid x_1, x_2) \geq \theta, \\
        0 & \text{otherwise}.
    \end{cases}
\end{equation}
}
\ec{In our analysis, we explore two scenarios: 1) The standard choice for a \textit{well-calibrated} model is to set $\theta=0.5$. This is what a typical user would choose, particularly given that the NOAA SWPC website does not provide any alternative information. 2) Alternatively, one can choose the threshold that maximizes specific metrics. Here, we choose to maximize True Skill Statistic (TSS). }

\subsection{Interpreting Metrics in Imbalanced Contexts}

\ai{The extreme imbalance between flare and no-flare days poses challenges for evaluating forecast skill.} \ec{\rev{For example, in our dataset of 9,828 days, M-class flares occurred only on $20.6\%$ of days and X-class flares only on $2.6\%$. As a result, a model that always predicts no flare achieves $79.4\%$ accuracy for M-class and $97.4\%$ for X-class.}} \ai{ Despite appearing high, this reflects a trivial solution and provides no actionable forecasting value,} \ec{hence the need to look at other metrics, such as Probability of Detection (POD, or Recall), Critical Success Index (CSI), and Heidke Skill Score (HSS) as key metrics in this study. The Brier score, which is often used to assess probabilistic forecast suffers the same shortfall: a model that always predicts no flare has a Brier score of 0.026 for X-class flares in our dataset  (Brier  = 0 is a perfect prediction).} 

\ai{High recall often comes at the cost of low precision, especially for rare events like X-class flares. We therefore also examine the False Alarm Ratio (FAR) and Precision to capture this tradeoff. While a high recall may be desirable for operational purposes (e.g., safety), a high false alarm rate can erode trust in forecasts.} \ec{It is worth noting  that in a highly imbalanced, mostly TN, dataset, FPs can be "hidden" by the huge number of TNs. In this scenario, TSS is compromised since FPR tends to 0 even though FPs are very high thus giving a misleading sense of high TSS. Many flare prediction papers fall prey to this and report relatively high TSS values (e.g., $>0.8$) from their model while failing to note that the FAR is extremely high, therefore making their model unsuitable for operational forecasting.}

\ai{Together, these baseline models and evaluation strategies provide a rigorous framework to assess the SWPC operational forecasts} \ec{ (or any other flare forecasting method with sufficient data for statistical analysis) using simple, interpretable baselines based on easily accessible solar activity features.}

\section{Results}
\subsection{Reliability of SWPC Probabilistic Forecasts}

\ai{We begin our analysis by evaluating the reliability of the SWPC forecasts using reliability diagrams, which compare the forecast probabilities with the observed frequency of events. A forecast is considered \textit{reliable} if, over all days when a certain probability $p$ is forecast, a flare occurs approximately $p$\% of the time.

Unlike many probabilistic forecast systems that issue probabilities with fine granularity, the SWPC forecasts are limited to a discrete set of probability values. Specifically, each lead time and flare class uses a fixed list of possible probabilities (percentage):}

\begin{itemize}
\ec{
    \item M-class forecasts (24h, 48h, 72h): \{1, 5, 10, 15, ..., 95\} (20 values)
    \item X-class 24h: \{1, 2, 5, 10, ..., 75\} (except for 65, 70 - 15 values)
    \item X-class 48h: \{1, 2, 5, 10, ..., 75\} (except for 65 - 16 values)
    \item X-class 72h: \{1, 2, 5, 10, ..., 75\} (except for 70 - 16 values)}
\end{itemize}

\ec{In passing, we notice that on the day 2007-09-04 the issued probabilities were 71/71/71 for M-class, which we considered a typo and changed to 70/70/70 (being the only one instance of 71\% probability in the whole database).}

\ai{As a result, we plot the observed relative frequencies directly against these discrete forecast values, rather than grouping into broader bins. Each point on the diagram corresponds to one of the forecast probability values, and its vertical position reflects the empirical probability of flare occurrence given that forecast value.} \ec{
Perfect reliability corresponds to the diagonal $y = x$ line, where the predicted probability equals the observed frequency. }

\begin{figure}[htbp]
    \centering
        \includegraphics[width=\linewidth]{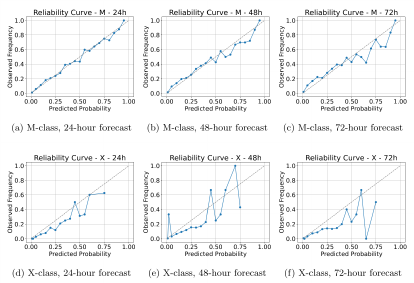}
    \caption{Reliability diagrams for M-class and X-class flare forecasts at 24-hour, 48-hour, and 72-hour lead times. The diagonal dashed line represents perfect reliability.}
    \label{fig:reliability-diagrams}
\end{figure}

\ec{
\textbf{M-class flare forecasts.}
The top row of Figure~\ref{fig:reliability-diagrams} shows the reliability diagrams for M-class forecasts at 24-hour, 48-hour, and 72-hour lead times. At 24-hour lead time, the forecasts are generally well-calibrated over the whole probability range. At longer lead times, forecast probabilities become less reliable, with a tendency toward overconfidence. For example, the 72-hour forecast at the predicted probability 80\% has an observed frequency of only 60\%.}

\ai{\textbf{X-class flare forecasts.}
The reliability diagrams for the X-class forecasts are shown in the bottom row of Figure~\ref{fig:reliability-diagrams}. Due to the rarity of X-class flares, the reliability curves exhibit greater variance and less smoothness, especially at higher probability bins where data are sparse. At all lead times, forecasts tend to be overconfident: the predicted probability exceeds the observed frequency of events, especially for probabilities above 20\%. This is consistent with a high FAR and is expected in rare-event forecasting when attempting to maximize recall.} \ec{The 48-hour forecast (middle column) shows an inconsistent behavior with the reliability curve crossing the diagonal line several times.}

\ai{The discrete nature of the forecast probabilities limits the resolution of the reliability curves, but also simplifies interpretation. Overall, the reliability analysis reveals that M-class forecasts are reasonably well-calibrated at short lead times, while X-class forecasts tend to be overconfident and degrade more quickly with increasing lead time. These patterns should be considered when interpreting the deterministic performance metrics presented in the next sections.}

\begin{table}
{\small
\caption{Comparison of flare prediction models across metrics M class, 24 hours ahead (Clim. = Climatology; Pers. = Persistence; NB = Naive Bayes; LR = Logistic Regression; BA = Baseline Average). Bold text indicates the best model for that metric.}
\label{tab:model_comparison_M_24}
\begin{tabular}{l|ccccccccccc}
\toprule
Model & Accuracy & Precision & Recall & F1 & Brier & AUC & CSI & POD & FAR & TSS & HSS \\
\toprule
SWPC & \textbf{0.84} & 0.62 & 0.53 & 0.57 & \textbf{0.11} & \textbf{0.87} & 0.40 & 0.53 & 0.38 & 0.44 & \textbf{0.47} \\
Clim. & 0.80 & 0.53 & 0.34 & 0.42 & 0.13 & 0.77 & 0.26 & 0.34 & 0.47 & 0.26 & 0.30 \\
Pers. & 0.82 & 0.57 & 0.57 & 0.57 & 0.18 & 0.73 & 0.40 & 0.57 & 0.43 & 0.46 & 0.46 \\
NB & 0.63 & 0.35 & \textbf{0.93} & 0.51 & 0.36 & 0.79 & 0.34 & \textbf{0.93} & 0.65 & \textbf{0.48} & 0.30 \\
LR & 0.81 & \textbf{0.64} & 0.15 & 0.24 & 0.13 & 0.83 & 0.13 & 0.15 & \textbf{0.36} & 0.12 & 0.17 \\
BA & 0.82 & 0.57 & 0.59 & \textbf{0.58} & 0.12 & 0.86 & \textbf{0.41} & 0.59 & 0.43 & 0.47 & \textbf{0.47} \\
\end{tabular}
}
\end{table}

\begin{table}
{\small
\caption{Comparison of flare prediction models across metrics M class, 48 hours ahead (Clim. = Climatology; Pers. = Persistence; NB = Naive Bayes; LR = Logistic Regression; BA = Baseline Average). Bold text indicates the best model for that metric.}
\label{tab:model_comparison_M_48}
\begin{tabular}{l|ccccccccccc}
\toprule
Model & Accuracy & Precision & Recall & F1 & Brier & AUC & CSI & POD & FAR & TSS & HSS \\
\toprule
SWPC & \textbf{0.82} & 0.57 & 0.46 & 0.51 & \textbf{0.12} & \textbf{0.85} & 0.34 & 0.46 & 0.43 & 0.37 & 0.40 \\
Clim. & 0.80 & 0.52 & 0.24 & 0.33 & 0.14 & 0.75 & 0.20 & 0.24 & 0.48 & 0.19 & 0.23 \\
Pers. & 0.81 & 0.53 & 0.53 & 0.53 & 0.19 & 0.70 & 0.36 & 0.53 & 0.47 & 0.41 & 0.41 \\
NB & 0.62 & 0.34 & \textbf{0.91} & 0.50 & 0.37 & 0.78 & 0.33 & \textbf{0.91} & 0.66 & \textbf{0.46} & 0.29 \\
LR & 0.81 & \textbf{0.64} & 0.13 & 0.22 & 0.13 & 0.82 & 0.12 & 0.13 & \textbf{0.36} & 0.11 & 0.16 \\
BA & 0.81 & 0.53 & 0.55 & \textbf{0.54} & 0.13 & 0.84 & \textbf{0.37} & 0.55 & 0.47 & 0.42 & \textbf{0.42} \\
\end{tabular}
}
\end{table}

\begin{table}
{\small
\caption{Comparison of flare prediction models across metrics M class, 72 hours ahead (Clim. = Climatology; Pers. = Persistence; NB = Naive Bayes; LR = Logistic Regression; BA = Baseline Average). Bold text indicates the best model for that metric.}
\label{tab:model_comparison_M_72}
\begin{tabular}{l|ccccccccccc}
\toprule
Model & Accuracy & Precision & Recall & F1 & Brier & AUC & CSI & POD & FAR & TSS & HSS \\
SWPC & \textbf{0.81} & 0.55 & 0.43 & 0.48 & \textbf{0.13} & \textbf{0.83} & 0.32 & 0.43 & 0.45 & 0.34 & 0.37 \\
Clim. & 0.79 & 0.49 & 0.21 & 0.29 & 0.14 & 0.74 & 0.17 & 0.21 & 0.51 & 0.15 & 0.19 \\
Pers. & 0.79 & 0.49 & 0.49 & 0.49 & 0.21 & 0.68 & 0.33 & 0.49 & 0.51 & 0.36 & 0.36 \\
NB & 0.62 & 0.34 & \textbf{0.90} & 0.49 & 0.38 & 0.77 & 0.33 & \textbf{0.90} & 0.66 & \textbf{0.45} & 0.28 \\
LR & \textbf{0.81} & \textbf{0.67} & 0.11 & 0.19 & \textbf{0.13} & 0.81 & 0.11 & 0.11 & \textbf{0.33} & 0.10 & 0.14 \\
BA & 0.80 & 0.50 & 0.51 & \textbf{0.51} & 0.14 & 0.82 & \textbf{0.34} & 0.51 & 0.50 & 0.38 & \textbf{0.38} \\
\end{tabular}
}
\end{table}

\subsection{Standard threshold: $\theta=0.5$}

\ai{Here, we evaluate flare prediction performance across multiple models and forecast lead times (24-, 48-, and 72-hours) for both M-class and X-class events, using a standard probability threshold of 0.5. Tables~\ref{tab:model_comparison_M_24}--\ref{tab:model_comparison_X_72} summarize the results. Overall, the NOAA SWPC forecast is consistently outperformed by baseline models across nearly all skill scores—most notably for metrics less sensitive to class imbalance, such as the F1, HSS, and CSI.}

\subsubsection{Performance on M-class Flare Forecasts}
\ai{
\rev{Across all lead times (24h, 48h, and 72h), the SWPC model achieves the highest accuracy and lowest Brier scores. However, these metrics can be misleading in highly imbalanced datasets like solar flare occurrence, where non-events dominate. High accuracy may reflect a tendency to predict "no flare" by default, and a low Brier score can indicate cautious probability estimates rather than true predictive skill.}

\rev{When focusing on event-focused metrics—such as F1 score, Critical Success Index (CSI), and Heidke Skill Score (HSS)—the Baseline Average (BA) model consistently outperforms SWPC at all lead times. At 24 hours, BA achieves the best F1 score (0.58), CSI (0.41), and matches SWPC in HSS (0.47), while offering higher recall (0.59) with comparable precision. Similar trends persist at 48 and 72 hours, where BA maintains stronger overall balance between detecting flares and avoiding false alarms.}

\rev{Remarkably, even the simple persistence model—using no training and based solely on the previous day's flare activity—performs on par with, or only marginally below, the SWPC forecast. At all lead times, persistence achieves similar or better scores than SWPC in recall, F1, CSI, and HSS.}
}
\subsubsection{X-class Flares}

\begin{table}
{\small
\caption{Comparison of flare prediction models across metrics X class, 24 hours ahead (Clim. = Climatology; Pers. = Persistence; NB = Naive Bayes; BA = Baseline Average). Bold text indicates the best model for that metric.}
\label{tab:model_comparison_X_24}
\begin{tabular}{l|ccccccccccc}
\toprule
Model & Accuracy & Precision & Recall & F1 & Brier & AUC & CSI & POD & FAR & TSS & HSS \\
\toprule
SWPC & \textbf{0.97} & \textbf{0.39} & 0.08 & 0.13 & \textbf{0.02} & \textbf{0.87} & 0.07 & 0.08 & \textbf{0.61} & 0.08 & 0.12 \\
Clim. & \textbf{0.97} & 0.10 & 0.03 & 0.04 & 0.03 & 0.60 & 0.02 & 0.03 & 0.90 & 0.02 & 0.03 \\
Pers. & 0.96 & 0.21 & 0.21 & \textbf{0.21} & 0.04 & 0.60 & \textbf{0.12} & 0.21 & 0.79 & 0.19 & \textbf{0.19} \\
NB & 0.57 & 0.05 & \textbf{0.84} & 0.09 & 0.43 & 0.74 & 0.05 & \textbf{0.84} & 0.95 & \textbf{0.40} & 0.04 \\
BA & 0.96 & 0.19 & 0.18 & 0.18 & 0.07 & 0.80 & 0.10 & 0.18 & 0.81 & 0.16 & 0.16 \\
\end{tabular}
}
\end{table}

\begin{table}
{\small
\caption{Comparison of flare prediction models across metrics X class, 48 hours ahead (Clim. = Climatology; Pers. = Persistence; NB = Naive Bayes; BA = Baseline Average). Bold text indicates the best model for that metric.}
\label{tab:model_comparison_X_48}
\begin{tabular}{l|ccccccccccc}
\toprule
Model & Accuracy & Precision & Recall & F1 & Brier & AUC & CSI & POD & FAR & TSS & HSS \\
SWPC & \textbf{0.97} & \textbf{0.33} & 0.06 & 0.10 & \textbf{0.02} & \textbf{0.84} & 0.05 & 0.06 & \textbf{0.67} & 0.06 & 0.09 \\
Clim. & \textbf{0.97} & 0.06 & 0.01 & 0.02 & 0.03 & 0.57 & 0.01 & 0.01 & 0.94 & 0.01 & 0.01 \\
Pers. & 0.96 & 0.20 & 0.20 & \textbf{0.20} & 0.04 & 0.59 & \textbf{0.11} & 0.20 & 0.80 & 0.18 & \textbf{0.18} \\
NB & 0.56 & 0.05 & \textbf{0.82} & 0.09 & 0.44 & 0.73 & 0.05 & \textbf{0.82} & 0.95 & \textbf{0.37} & 0.04 \\
BA & 0.96 & 0.19 & 0.18 & 0.19 & 0.07 & 0.77 & 0.10 & 0.18 & 0.81 & 0.16 & 0.16 \\
\end{tabular}
}
\end{table}

\begin{table}
{\small
\caption{Comparison of flare prediction models across metrics X class, 72 hours ahead (Clim. = Climatology; Pers. = Persistence; NB = Naive Bayes; BA = Baseline Average). Bold text indicates the best model for that metric.}
\label{tab:model_comparison_X_72}
\begin{tabular}{l|ccccccccccc}
\toprule
Model & Accuracy & Precision & Recall & F1 & Brier & AUC & CSI & POD & FAR & TSS & HSS \\
SWPC & \textbf{0.97} & \textbf{0.29} & 0.05 & 0.08 & \textbf{0.02} & \textbf{0.81} & 0.04 & 0.05 & \textbf{0.71} & 0.04 & 0.07 \\
Clim. & \textbf{0.97} & 0.06 & 0.01 & 0.02 & 0.03 & 0.53 & 0.01 & 0.01 & 0.94 & 0.01 & 0.01 \\
Pers. & 0.96 & 0.19 & 0.19 & \textbf{0.19} & 0.04 & 0.58 & \textbf{0.10} & 0.19 & 0.81 & 0.16 & \textbf{0.16} \\
NB & 0.55 & 0.04 & \textbf{0.79} & 0.08 & 0.44 & 0.72 & 0.04 & \textbf{0.79} & 0.96 & \textbf{0.34} & 0.04 \\
BA & 0.96 & 0.18 & 0.17 & 0.18 & 0.07 & 0.73 & \textbf{0.10} & 0.17 & 0.82 & 0.15 & \textbf{0.16} \\
\end{tabular}
}
\end{table}
\ai{
The performance gap is even more pronounced for the rare but operationally critical X-class flares. The SWPC forecast exhibits low recall ($\leq0.08$) and corresponding CSI values ($\leq0.07$) at all forecast horizons, indicating poor detection rates. 

\rev{Across all lead times (24h, 48h, 72h), the SWPC model achieves the highest accuracy (0.97) and the lowest Brier scores (0.02), and maintains the best AUC values (up to 0.87). However, as with M-class flares, these metrics can be misleading in highly imbalanced datasets. The SWPC model exhibits very low recall (as low as 0.05–0.08) and F1 scores (0.08–0.13), meaning it fails to capture most X-class flare events. The result is a model that rarely predicts flares and is therefore of limited use for event detection.}

\rev{In contrast, the persistence model achieves significantly better balance in key operational metrics. It consistently outperforms SWPC in recall (0.19–0.21), F1 (up to 0.21), Critical Success Index (CSI), and both TSS and HSS, despite having no underlying data model. For example, at 24h lead time, persistence achieves the highest CSI (0.12) and HSS (0.19), highlighting its effectiveness as a zero-cost benchmark.}
}
\rev{\ai{The Baseline Average (BA), constructed here by averaging the predictions of climatology, persistence, and Naive Bayes, consistently performs more robustly than SWPC} \ec{ (logistic regression is excluded in this case for its low performance).} \ai{While it does not lead in any individual metric, it achieves a more balanced compromise between precision and recall, yielding higher F1 scores and CSI values than climatology or SWPC at all \tb{lead times}. Notably, at 72h, BA matches persistence in HSS (0.16), with a similar CSI (0.10) and recall (0.17), and outperforms SWPC in F1, CSI, and HSS. These results reinforce the idea that even simple ensemble strategies can outperform or rival complex, opaque forecasting systems in rare event prediction.}}

\ai{
\rev{Overall, the SWPC forecast for X-class flares shows poor event detection skill and is routinely outperformed by persistence and ensemble baselines. This underscores the need for operational models to prioritize recall and discrimination when dealing with rare but high-impact events.}
}

\subsubsection{Summary of forecast comparisons}
\ai{
\rev{The comprehensive validation of flare forecasts across M-class and X-class events, and for lead times of 24, 48, and 72 hours, reveals consistent patterns in model performance. For M-class flares, the SWPC forecast achieves the highest accuracy and Brier scores, but these metrics are inflated by class imbalance and do not reflect true event detection skill. In contrast, the Baseline Average (BA) model—a simple ensemble of zero-cost models—consistently achieves superior scores in event-sensitive metrics such as F1, CSI, TSS, and HSS, indicating better balance between sensitivity and specificity. Notably, even the persistence model, which requires no training and uses no flare history beyond the previous day, performs on par with or slightly below SWPC in most skill scores.}

\rev{For X-class flares, the limitations of the SWPC forecast become even more pronounced. While it maintains the highest accuracy and lowest Brier scores, its recall is very low and it fails to detect most flare events, resulting in low F1 and CSI scores. Persistence again outperforms SWPC in almost all event-relevant metrics, while Naive Bayes shows high recall but unacceptably high false alarm rates. The Baseline Average model offers a well-balanced alternative, outperforming SWPC in key scores like F1, CSI, and HSS at all horizons. Logistic regression was excluded from the X-class comparison due to poor calibration and negligible recall.}

\rev{Overall, the findings suggest that SWPC forecasts do not significantly outperform zero-cost or simple statistical baselines, particularly in rare-event detection.} 
}
\\

\ec{
It is important to stress that comparisons with zero-cost baseline models are not meant to suggest these models as viable forecasting alternatives on their own, but rather that they should be considered as additional tools for human forecasters when issuing 1--3 day flare forecasts. Ideally, an official forecast should be developed as a multi-model ensemble with human-in-the-loop interpretation that takes into account the skill of experienced space weather forecasters developed over many years of observing solar behavior.}

\subsection{Optimal probability threshold}
\ec{In this section, we determine the probability threshold $\theta$ that optimizes the TSS metric for each model independently. The optimal values of $\theta$ are summarized in Table~\ref{tab:optimal_theta}. We recall that this value sets the threshold for positive/negatives for the binary classification metrics. Typically, a well-calibrated model would have an optimal threshold close to 0.5.} 

\ai{The optimal probability thresholds presented in Table~\ref{tab:optimal_theta} highlight substantial calibration issues with the SWPC forecasts, particularly for X-class flares, where thresholds as low as 0.05 are required to produce any positive forecasts. These extremely low thresholds indicate that SWPC X-class flare forecasts are systematically under-confident, necessitating aggressive post-processing to extract meaningful predictions. In contrast, the logistic regression model exhibits more moderate and consistent thresholds, suggesting better calibration} \ec{(the threshold for persistence is either zero or one by construction).}

\ai{Applying optimized probability thresholds for binary classification improves the event detection capabilities of all models, leading to a more meaningful evaluation of forecast skill under imbalanced conditions. The results for M-class flares at 24, 48, and 72 hours (Tables~\ref{tab:model_comparison_M_24_opt_thr}–\ref{tab:model_comparison_M_72_opt_thr}) reveal that the SWPC model improves its recall substantially (e.g., 0.86 at 24h) while maintaining a low Brier score and high AUC, suggesting good calibration and discrimination. However, when judged by event-focused metrics such as F1, CSI, TSS, and HSS, the performance gap between SWPC and baseline models narrows considerably—and in some cases, disappears.

Performance comparisons across all forecast lead times and flare classes (Tables~\ref{tab:model_comparison_M_24_opt_thr} to~\ref{tab:model_comparison_X_72_opt_thr}) consistently show that baseline models—particularly climatology and persistence—either match or outperform the SWPC model on most metrics. While SWPC tends to achieve high recall values (up to 0.93 for X-class flares at 24 hours; Table~\ref{tab:model_comparison_X_24_opt_thr}), this comes at the cost of extremely high false alarm rates (FARs), often exceeding 90\%. These inflated FARs greatly reduce the practical utility of the SWPC forecasts.

Persistence, while simplistic, achieves the highest accuracy and precision in many settings (e.g., X-class, all lead times; Tables~\ref{tab:model_comparison_X_24_opt_thr}--\ref{tab:model_comparison_X_72_opt_thr}), with much lower FARs than SWPC, despite its lower recall. 

The poor calibration and elevated FARs associated with the SWPC model suggest that the forecasts, in their current form, are not well-suited for operational decision-making without significant recalibration or supplementation with baseline or statistical methods.

Overall, the use of optimized thresholds reveals that SWPC's flare forecast is closely matched or exceeded by simple statistical models and the baseline average. The persistence model, in particular, continues to provide competitive skill with no learning or parameter tuning, emphasizing the need for operational models to demonstrate clear added value over trivial heuristics.}

\subsection{Best Trade-Off Across Metrics}
\ai{
\rev{When considering all metrics jointly—particularly those most relevant for operational performance such as F1 score, Critical Success Index (CSI), True Skill Statistic (TSS), and Heidke Skill Score (HSS)— SWPC forecasts are not consistently outperforming the \textbf{Baseline Average (BA)} for M-class flares and \textbf{Persistence} for X-class flares.}

\rev{For M-class flare prediction, the Baseline Average consistently achieves top or near-top performance across F1, CSI, and HSS for all forecast horizons (24h, 48h, and 72h). It combines high recall with moderate false alarm rates and outperforms or matches the SWPC forecast in all threshold-dependent metrics. The ensemble nature of BA allows it to capture the strengths of its component models (climatology, Naive Bayes, and logistic regression), leading to more balanced and reliable classification performance.}

\rev{For X-class flares, Persistence emerges as the most reliable and balanced model. While the SWPC forecast achieves extremely high recall, it does so at the expense of precision, resulting in very high false alarm rates and low F1 and CSI scores. In contrast, Persistence offers moderate recall (0.19–0.21), substantially higher precision (0.19–0.21), and the best F1, CSI, and HSS values across all lead times. It achieves this without any learning or model complexity, highlighting the importance of evaluating against strong baseline methods.}
}

\begin{table}
\caption{Optimal values of probability threshold $\theta$, \rev{defined as the value that maximizes TSS.}}
\label{tab:optimal_theta}
\begin{tabular}{l|cccccc}
\toprule
Model & M 24h & M 48h & M 72h & X 24h & X 48h & X 72h\\
\toprule
SWPC & 0.20 & 0.20 & 0.15 & 0.05& 0.05& 0.05\\ 
Climatology & 0.15 & 0.14 & 0.16 & 0.03& 0.03& 0.03\\
Persistence & 1.00 & 1.00 & 1.00 & 1.00& 1.00& 1.00\\
Naive Bayes & 1.00 & 1.00 & 1.00 & 1.00& 1.00& 1.00\\
Logistic Regression & 0.18 & 0.18 & 0.17 & 0.03 & 0.02& 0.03 \\
Baseline Average & 0.36 & 0.36 & 0.34 & 0.26 & 0.26 & 0.26\\
\end{tabular}
\end{table}

\begin{table}
{\small
\caption{Comparison of flare prediction models across metrics M class, 24 hours ahead (optimized threshold). (Clim. = Climatology; Pers. = Persistence; NB = Naive Bayes; BA = Baseline Average). Bold text indicates the best model for that metric.}
\label{tab:model_comparison_M_24_opt_thr}
\begin{tabular}{lllccccccccc}
\toprule
Model & Accuracy & Precision & Recall & F1 & Brier & AUC & CSI & POD & FAR & TSS & HSS \\
\midrule
SWPC & 0.75 & 0.44 & 0.86 & \textbf{0.58} & \textbf{0.11} & \textbf{0.87} & \textbf{0.41} & 0.86 & 0.56 & \textbf{0.58} & 0.43 \\
Clim. & 0.72 & 0.41 & 0.80 & 0.54 & 0.13 & 0.77 & 0.37 & 0.80 & 0.59 & 0.50 & 0.37 \\
Pers. & \textbf{0.82} & \textbf{0.57} & 0.57 & 0.57 & 0.18 & 0.73 & 0.40 & 0.57 & \textbf{0.43} & 0.46 & \textbf{0.46} \\
NB & 0.67 & 0.37 & \textbf{0.89} & 0.52 & 0.36 & 0.79 & 0.35 & \textbf{0.89} & 0.63 & 0.50 & 0.33 \\
LR & 0.72 & 0.42 & 0.85 & 0.56 & 0.13 & 0.83 & 0.39 & 0.85 & 0.58 & 0.54 & 0.39 \\
BA & 0.76 & 0.46 & 0.81 & \textbf{0.58} & 0.12 & 0.86 & \textbf{0.41} & 0.81 & 0.54 & 0.56 & 0.43 \\
\end{tabular}
}
\end{table}

\begin{table}
{\small
\caption{Comparison of flare prediction models across metrics M class, 48 hours ahead (optimized threshold). (Clim. = Climatology; Pers. = Persistence; NB = Naive Bayes; BA = Baseline Average). Bold text indicates the best model for that metric.}
\label{tab:model_comparison_M_48_opt_thr}
\begin{tabular}{lllccccccccc}
\toprule
Model & Accuracy & Precision & Recall & F1 & Brier & AUC & CSI & POD & FAR & TSS & HSS \\
\midrule
SWPC & 0.74 & 0.43 & 0.83 & \textbf{0.56} & \textbf{0.12} & \textbf{0.85} & \textbf{0.39} & 0.83 & 0.57 & \textbf{0.54} & 0.40 \\
Clim. & 0.69 & 0.38 & 0.80 & 0.51 & 0.14 & 0.75 & 0.35 & 0.80 & 0.62 & 0.46 & 0.33 \\
Pers. & \textbf{0.81} & \textbf{0.53} & 0.53 & 0.53 & 0.19 & 0.70 & 0.36 & 0.53 & \textbf{0.47} & 0.41 & \textbf{0.41} \\
NB & 0.66 & 0.36 & \textbf{0.87} & 0.51 & 0.37 & 0.78 & 0.34 & \textbf{0.87} & 0.64 & 0.47 & 0.31 \\
LR & 0.71 & 0.40 & 0.83 & 0.54 & 0.13 & 0.82 & 0.37 & 0.83 & 0.60 & 0.51 & 0.37 \\
BA & 0.75 & 0.44 & 0.78 & \textbf{0.56} & 0.13 & 0.84 & \textbf{0.39} & 0.78 & 0.56 & 0.52 & 0.40 \\
\end{tabular}
}
\end{table}

\begin{table}
{\small
\caption{Comparison of flare prediction models across metrics M class, 72 hours ahead (optimized threshold). (Clim. = Climatology; Pers. = Persistence; NB = Naive Bayes; BA = Baseline Average). Bold text indicates the best model for that metric.}
\label{tab:model_comparison_M_72_opt_thr}
\begin{tabular}{lllccccccccc}
\toprule
Model & Accuracy & Precision & Recall & F1 & Brier & AUC & CSI & POD & FAR & TSS & HSS \\
\midrule
SWPC & 0.71 & 0.40 & 0.84 & \textbf{0.54} & \textbf{0.13} & \textbf{0.83} & \textbf{0.37} & 0.84 & 0.60 & \textbf{0.51} & \textbf{0.36} \\
Clim. & 0.69 & 0.37 & 0.76 & 0.50 & 0.14 & 0.74 & 0.33 & 0.76 & 0.63 & 0.43 & 0.31 \\
Pers. & \textbf{0.79} & \textbf{0.49} & 0.49 & 0.49 & 0.21 & 0.68 & 0.33 & 0.49 & \textbf{0.51} & 0.36 & \textbf{0.36} \\
NB & 0.65 & 0.36 & \textbf{0.86} & 0.50 & 0.38 & 0.77 & 0.34 & \textbf{0.86} & 0.64 & 0.46 & 0.30 \\
LR & 0.69 & 0.39 & 0.84 & 0.53 & \textbf{0.13} & 0.81 & 0.36 & 0.84 & 0.61 & 0.50 & 0.35 \\
BA & 0.71 & 0.40 & 0.79 & 0.53 & 0.14 & 0.82 & 0.36 & 0.79 & 0.60 & 0.48 & 0.35 \\\end{tabular}
}
\end{table}

\begin{table}
{\small
\caption{Comparison of flare prediction models across metrics X class, 24 hours ahead (optimized threshold). (Clim. = Climatology; Pers. = Persistence; NB = Naive Bayes; BA = Baseline Average). Bold text indicates the best model for that metric.}
\label{tab:model_comparison_X_24_opt_thr}
\begin{tabular}{lllccccccccc}
\toprule
Model & Accuracy & Precision & Recall & F1 & Brier & AUC & CSI & POD & FAR & TSS & HSS \\
\midrule
SWPC & 0.69 & 0.07 & \textbf{0.93} & 0.14 & \textbf{0.02} & \textbf{0.87} & 0.07 & \textbf{0.93} & 0.93 & \textbf{0.61} & 0.09 \\
Clim. & 0.88 & 0.09 & 0.38 & 0.14 & 0.03 & 0.60 & 0.08 & 0.38 & 0.91 & 0.27 & 0.10 \\
Pers. & \textbf{0.96} & \textbf{0.21} & 0.21 & \textbf{0.21} & 0.04 & 0.60 & \textbf{0.12} & 0.21 & \textbf{0.79} & 0.19 & \textbf{0.19} \\
NB & 0.62 & 0.05 & 0.80 & 0.10 & 0.43 & 0.74 & 0.05 & 0.80 & 0.95 & 0.42 & 0.05 \\
LR & 0.72 & 0.06 & 0.69 & 0.11 & \textbf{0.02} & 0.77 & 0.06 & 0.69 & 0.94 & 0.42 & 0.07 \\
BA & 0.61 & 0.05 & 0.84 & 0.10 & 0.05 & 0.80 & 0.05 & 0.84 & 0.95 & 0.45 & 0.06 \\\end{tabular}
}
\end{table}

\begin{table}
{\small
\caption{Comparison of flare prediction models across metrics X class, 48 hours ahead (optimized threshold). (Clim. = Climatology; Pers. = Persistence; NB = Naive Bayes; BA = Baseline Average). Bold text indicates the best model for that metric.}
\label{tab:model_comparison_X_48_opt_thr}
\begin{tabular}{lllccccccccc}
\toprule
Model & Accuracy & Precision & Recall & F1 & Brier & AUC & CSI & POD & FAR & TSS & HSS \\
\midrule
SWPC & 0.70 & 0.07 & \textbf{0.88} & 0.13 & \textbf{0.02} & \textbf{0.84} & 0.07 & \textbf{0.88} & 0.93 & \textbf{0.58} & 0.09 \\
Clim. & 0.80 & 0.05 & 0.41 & 0.10 & 0.03 & 0.57 & 0.05 & 0.41 & 0.95 & 0.22 & 0.05 \\
Pers. & \textbf{0.96} & \textbf{0.20} & 0.20 & \textbf{0.20} & 0.04 & 0.59 & \textbf{0.11} & 0.20 & \textbf{0.80} & 0.18 & \textbf{0.18} \\
NB & 0.64 & 0.05 & 0.75 & 0.10 & 0.44 & 0.73 & 0.05 & 0.75 & 0.95 & 0.38 & 0.05 \\
LR & 0.57 & 0.05 & 0.84 & 0.09 & \textbf{0.02} & 0.76 & 0.05 & 0.84 & 0.95 & 0.40 & 0.04 \\
BA & 0.62 & 0.05 & 0.80 & 0.10 & 0.05 & 0.78 & 0.05 & 0.80 & 0.95 & 0.41 & 0.05 \\
\end{tabular}
}
\end{table}

\begin{table}
{\small
\caption{Comparison of flare prediction models across metrics X class, 72 hours ahead (optimized threshold). (Clim. = Climatology; Pers. = Persistence; NB = Naive Bayes; BA = Baseline Average). Bold text indicates the best model for that metric.}
\label{tab:model_comparison_X_72_opt_thr}
\begin{tabular}{lllccccccccc}
\toprule
Model & Accuracy & Precision & Recall & F1 & Brier & AUC & CSI & POD & FAR & TSS & HSS \\
\midrule
SWPC & 0.71 & 0.07 & \textbf{0.83} & 0.13 & \textbf{0.02} & \textbf{0.81} & 0.07 & \textbf{0.83} & 0.93 & \textbf{0.53} & 0.08 \\
Clim. & 0.80 & 0.04 & 0.32 & 0.08 & 0.03 & 0.53 & 0.04 & 0.32 & 0.96 & 0.13 & 0.03 \\
Pers. & \textbf{0.96} & \textbf{0.19} & 0.19 & \textbf{0.19} & 0.04 & 0.58 & \textbf{0.10} & 0.19 & \textbf{0.81} & 0.16 & \textbf{0.16} \\
NB & 0.63 & 0.05 & 0.73 & 0.09 & 0.44 & 0.72 & 0.05 & 0.73 & 0.95 & 0.36 & 0.05 \\
LR & 0.67 & 0.05 & 0.72 & 0.10 & \textbf{0.02} & 0.75 & 0.05 & 0.72 & 0.95 & 0.38 & 0.05 \\
BA & 0.62 & 0.05 & 0.75 & 0.09 & 0.05 & 0.75 & 0.05 & 0.75 & 0.95 & 0.36 & 0.05 \\
\end{tabular}
}
\end{table}

\ai{
\subsection{Storm After the Calm: Forecasting the First Event After Extended Quiet}

A particularly challenging scenario for flare forecasting is the detection of a sudden, isolated X-class flare following a prolonged period of solar quiet. This situation, which we term the \textit{storm after the calm}, is especially relevant for human spaceflight, where early warning is critical to ensure astronauts can seek shelter in time.} \ec{To assess this, we focus on a particularly demanding scenario: days for which the number of prior flare-free days is larger than 30, meaning the Sun has been quiet for over a month. In this regime, we evaluate whether the model can detect the first sudden X-class flare after a prolonged period of inactivity.} \ai{The subset contains 66 positive cases (X-flares) and 6992 negatives, making it both operationally relevant and extremely imbalanced.}

\begin{figure}
\includegraphics{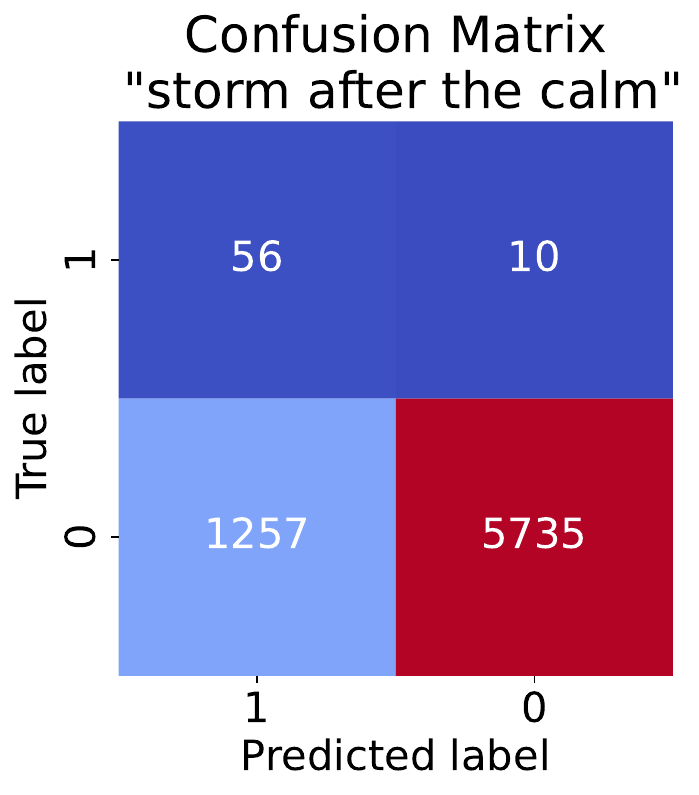}
\caption{Confusion matrix for 'storm after the calm' scenario (X flares)}\label{fig:stormcalm}
\end{figure}

\ai{Using the SWPC model’s optimal threshold for X-class flares at 24 hours (5\%), the confusion matrix is shown in Figure \ref{fig:stormcalm} (TP = 56, FN = 10, TN = 5735, FP = 1257).}

\ec{The results are concerning. Ten of the X-class flares were missed — 15\% or 1 in every 7 dangerous events. Each of these false negatives represents a potentially catastrophic failure. In human spaceflight, one missed X-flare can be lethal. A model that fails to anticipate such events, even occasionally, cannot be trusted in high-stakes operational contexts.}

\ai{Moreover, the SWPC forecast produces a large number of false alarms: 1,257 quiet days incorrectly predicted as flare-active out of 1,313 predicted positive} \ec{(95\% false alarm ratio), undermining its credibility and potentially leading to mission planners choosing to ignore the forecast altogether.}

\ai{
\subsection{All-Clear Events: Forecasting a Return to Quiet After X-Class Flares}

Complementary to the ``storm after the calm'' scenario, we evaluate the ability of the SWPC forecast to identify \textit{all-clear} periods following an X-class flare. Specifically, we define an all-clear event as a day with no X-class flares occurring \textbf{+1, +2, or +3 days} after an X-class flare. For each of these days, we assess the model's 24-hour forecast issued the day before.}

\ec{This is particularly crucial for human spaceflight \cite{ji2020all, sadykov2021all}. Planning for Extravehicular Activities (EVAs) or sorties on the surface of the Moon or Mars, during which astronauts have a greatly enhanced risk of radiation exposure due to the lack of spacecraft or habitat shielding, relies on identifying ``flare-free'' periods days in advance. In these activities, a false negative, or missed, X-class flare could result in lethal radiation exposure to the astronauts on the EVA. Therefore, the ability to accurately and reliably predict extended quiet periods, and to forecast the first sign of dangerous activity, is a defining test of a model’s usefulness.}

\ai{We construct a subset of forecast instances by selecting all days that are +1, +2, or +3 after a confirmed X-class flare, yielding a dataset focused specifically on the period immediately following high activity. For each instance, we apply the SWPC 24-hour forecast with its optimal threshold for X-class events (5\%) to determine whether the model predicts renewed flare activity.}

\begin{figure}
\includegraphics{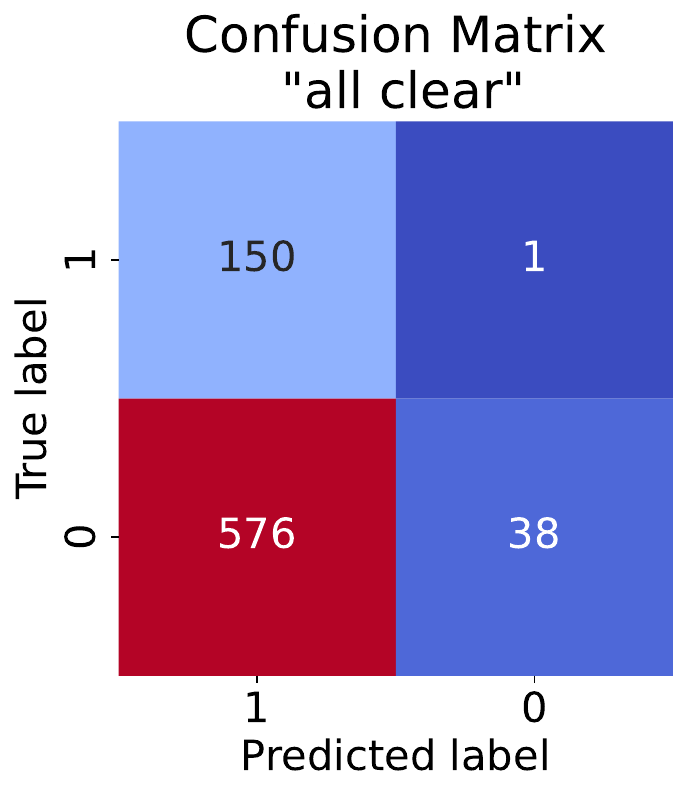}
\caption{Confusion matrix for 'all clear' scenario (X flares)}\label{fig:allclear}
\end{figure}

\ec{The confusion matrix for \textit{all-clear} periods is shown in Figure \ref{fig:allclear} (TP = 150, FN = 1, TN = 38, FP = 576).}

\ai{The forecast successfully identifies nearly all flare events (recall = 0.99), but it performs very poorly at identifying true all-clear days (precision = 0.21, FAR = 0.79). Of the 614 non-flaring days, only 38 are correctly predicted as quiet. The overwhelming number of false positives—576 out of 614—indicates a strong bias toward over-warning, severely limiting the forecast's value as a tool for operational confidence. }

\ai{In summary, although the SWPC forecast performs moderately well in aggregate statistics, its behavior in both all-clear and storm-after-the-calm scenarios reveals a fundamental limitation: it is not yet reliable enough} \ec{to be used in actionable decisions regarding human spaceflight safety. NASA is currently funding a center of excellence for developing new all-clear radiation event forecasting models \cite{zhao2023clear}. It will be imperative to compare the model(s) developed by this center objectively using the methods and metric described here (as well as others) to demonstrate improved forecast efficacy for the critical all-clear condition. }

\section{Conclusions}

\ec{This study presents a comprehensive verification of the NOAA/SWPC operational forecasts for M-class and X-class solar flares over the period 1998–2024. We find that SWPC forecasts perform comparably to, or are outperformed by, simple statistical baselines—including persistence, climatology, and lightweight machine learning classifiers such as logistic regression and Naive Bayes—across a wide range of skill metrics.} \ai{In many cases, particularly for X-class flares, SWPC forecasts demonstrate poor calibration, low precision, and high false alarm rates,} \ec{emphasizing the need to move beyond a single model or method for the issuance of official solar flare forecasts. }

\ec{The findings shown here are concerning in high-risk/high-impact operational contexts.} \ai{Of particular concern is the model's performance in ``all-clear" scenarios, where its inability to reliably detect the onset of dangerous X-class flares after prolonged quiet periods could lead to} \ec{life-threatening outcomes for astronauts executing EVAs or in unsheltered surface exploration missions on the Moon or Mars.}

\ec{We suggest that in addition to the current reliance solely on the McIntosh classification-based method, SWPC and other operational space weather forecasting offices should adopt a multi-model approach, using some or all of the comparison models shown here in an ensemble approach that produces a more reliably skilled flare forecast. In particular, operational forecasting offices should work towards transitioning both the simple baselines shown here as well as more advanced state-of-the-art ML flare prediction models \emph{once such models have been demonstrated to produce verifiably accurate and reliable predictions of solar flares.}

In reference to this latter condition, any solar flare forecasting or all-clear prediction models developed in a research setting should perform the type of basic forecast verification study shown here -- using the same baseline models and metrics for comparison -- \emph{before} being claimed as an advance over current methods. Many recent models developed for solar flare prediction have been published with, in some cases, $\mathrm{TSS} > 0.9$ and yet without reporting of the FAR, FPR, or other relevant ``negative metrics'' over a sufficiently long period to provide a credible evaluation of skill. While we do not wish to single out particular models for criticism, it is worth emphasizing that in highly imbalanced event classes like solar flares, $\mathrm{TN\ is\ often}\gg\mathrm{TP\ or\  FP}$ and thus the FPR is effectively zero, the TSS is artificially high (essentially just Recall), and extremely high FARs are effectively ignored. Indeed, specific model architectures have been developed to address the ``disguised FPR'' problem in solar flare prediction \cite{deshmukh_decreasing_2022}.

While meta-studies like \citeA{leka2019comparison} are useful for comparing the skills of a fixed set of models for a set period of solar activity, it should be the responsibility of the model developers themselves, or perhaps a neutral third-party such as NASA's Community Coordinated Modeling Center (CCMC), to run extensive and standardized forecast verification studies like the one demonstrated here prior to publishing a new prediction model and claiming superior predictive performance. 

We emphasize that forecast verification and model validation are separate and very different activities. Model validation is usually a ``nowcast verification'' where input to a model at a given time produces model state output that is compared to observations at that same time, i.e., with no forecast of the future system state. Model validation exercises often include extensive pre-processing of input data, pre-selection of favorable conditions and/or elimination of unfavorable conditions, and model tuning between runs to optimize performance on a given dataset, making such exercises essentially irrelevant to actual operational forecasting use. It is possible (and apparently not uncommon) for a space weather model to validate well and yet demonstrate low forecasting skill. 


In closing, we encourage all operational forecasting offices to (i) transition both baselines and state-of-the-art machine learning models, which have demonstrated robust performance and can be implemented at low computational cost, to operations as soon as practicable to supplement existing forecasting methods, and (ii) routinely perform and publish rigorous operational forecast verifications using open, reproducible methods and transparent metrics. Only by doing so can we build trust among the space weather end-user community and ensure that operational space weather products meet the demands of critical infrastructure applications.
}

\section*{Open Research Section}

The daily solar flare probability forecasts from NOAA/SWPC used in this study are publicly available at \url{ftp://ftp.swpc.noaa.gov/pub/warehouse/}. The solar flare event data prior to 2002 are available from \url{ftp://ftp.swpc.noaa.gov/pub/indices/events/}, while post-2002 data are accessed through the ASR flare catalogue from \url{https://github.com/helio-unitov/ASR_cat/releases/download/v1.1/f_1995_2024.csv}. All data used span the years 1996–2024 and are fully open access. A copy of all the data used in the paper, along with 
the code used for data preprocessing, analysis, and figure generation—including annotated Jupyter notebooks—is available on Zenodo at \url{https://doi.org/10.5281/zenodo.16620803} \cite{camporeale2025zenodo} and in a public GitHub repository at \url{https://github.com/ML-Space-Weather/solar_flare_verification}. This ensures the full reproducibility of the results presented in this paper.

\acknowledgments
This work was partly inspired by a conversation between one of the authors and a representative from the Dutch Ministry of Defense, who shared that the NOAA/SWPC flare forecast was routinely used under the assumption that it was a regularly verified and skillful operational product. A subsequent literature search motivated the study published here.

This work was partially supported by NASA under awards No 80NSSC23M0192, 80NSSC20K1580, 80NSSC21K1555.

\appendix
\section*{Appendix: On the Use of Generative AI in this Work}

Both authors are enthusiastic supporters of the responsible use of artificial intelligence in research and publishing.  We recognize both the transformative potential and the ethical challenges that generative AI tools pose to academic work. As scientists and as members of the editorial community at AGU, we believe that thoughtful experimentation with these technologies is essential to understanding their capabilities and limitations \cite{camporeale2024our}.

This manuscript provided a timely opportunity to systematically explore the extent to which generative AI—specifically, OpenAI's ChatGPT—can assist in writing a scientific paper. The topic of this study is particularly well-suited for such an experiment: it does not introduce any novel methodology but rather applies well-established forecast verification techniques to publicly available data. The primary novelty of the work lies in the breadth of the validation effort—spanning 26 years of solar flare forecasts—which, to our knowledge, has not been previously conducted or benchmarked against standard baselines at this scale.

The experiment was conducted as follows: the first draft of the manuscript was authored by EC, who deliberately used ChatGPT (free version) as extensively as possible to generate the initial text. We estimate that approximately 80\% of the initial draft was AI-generated or lightly edited by a human. The second author, TB, then revised and edited the text to ensure scientific accuracy, clarity, and coherence, without knowledge of which sections were generated by AI versus human-written. The resulting manuscript was then submitted for peer review in its final, collaboratively refined form. A similar exercise was then followed in addressing the reviewers' comments and preparing a revised version of the manuscript.

Our intent is to fully communicate to the journal readership which paragraphs have been AI-generated, so that everyone can judge by themselves the capability and limitations of generative AI in academic writing. In this experiment, the drafting and editing process was likely accelerated by a factor of two to three. Much of the accompanying software was also AI-generated and subsequently refined through human debugging and improvement. We found this especially valuable during the non particularly exciting phases of the project, such as data downloading, parsing, and cleaning.

The paper follows a \textit{Neverending Story} style \cite{ende1993neverending}. The text in \ai{black} has been AI-generated or only slightly edited (no more than a few words changed within a paragraph). The text in \ec{blue} has been either entirely human-generated or severely edited starting from an AI text. 

To promote transparency and reproducibility, we report in the Supplementary Information a list of prompts used during our chat with ChatGPT.

%
 \bibliography{biblio} 
 


%
%
%
%
%

\newpage
\section*{Supplementary Information}

As explained in the main text, here we report the prompts used in chatGPT to draft the manuscript. Note that many quantitative information (lists, tables, etc.) is copied and pasted from code or latex. We report here prompts without correcting for typos (LLMs are very good at understanding misspelled text).

\noindent\textbf{List of ChatGPT prompts used to draft the manuscript (first version):}

\begin{itemize}
\item This is a paper for the AGU journal Space Weather (or JGR: Machine Learning and Computation, still undecided). The objective is to validate the predictions made daily by the NOAA Space Weather Prediction Center (SWPC) for M-class and X-class flares. SWPC issues proabilisitc forecasts for 1,2,3 days ahead.
The introduction part will need to broadly introduce the importance of flares forecasting for space weather; explain how flares are classified in different classes; introduce existing flares catalogues.
 Mention the importance of catalogues for validation and training machine learning models. Mention that SWPC does not explicitly states how the forecasts are produced and, to the best of the authors knowdledge, has never completely validated its flares prediction.
Write all text (and subsequent) in latex (without the preambles). The tone is scientific writing. Cite relevant references.
\item Move on to the Data section.
We used data downloaded from ftp.swpc.noaa.gov/pub/warehouse/ that contains the daily prediction from 1996 to current in the form of text files. We preprocessed the data to parse the predictions to a single csv file. We also used the ASR catalogue, which contains flares in the period 2002-2024. The ASR catalogue contains also the official NOAA flare catalogue. For years prior to 2002 we have downloaded the official NOAA event reports from here: ftp://ftp.swpc.noaa.gov/pub/indices/events/ 
In total, the validation is done on a total of 10323 days. For each day a 'ground truth' label M-class and X-class is defined as a binary label (1 if a flare occurred anytime during that day, 0 otherwise).
Mention and describe the following plots to be included: number of flares vs day of the year; likelihood of a flare conditioned on how many prior days have gone without a flare (conditional probability). table with total number of positives and negatives (per flare class) and calculate the imbalance. 
\item add a paragraph about a plot of number of positive days over a 27-days period vs time,, which clearly shows the solar cycles.
Comment on the conditional probability of a flare occurring as a function of flare-free days
Especially for M-flares, there is more than 50\% probability that a M-flare is occurring, if it occurred the day before, while that is 20\% for X-class flares. After about 10 days there seems to be little dependence on the past number of flare-free days.
\item let's move to methodology.
Introduce the following metrics for classification: Accuracy': acc,
        'Precision': prec,
        'Recall': rec,
        'F1': f1,
        'Brier': brier,
'CSI': csi,
        'POD': pod, FAR, BIAS, TSS, HSS.
Explain the ROC curve and the concept of optimal threshold (that maximizes TSS). Explain that because the predicitons are probabilistic, we need to choose a threshold to use metrics based on deterministic binary classification.
introduce the idea of baseline models to compare with. We will use persistence (use ground truth from today as prediciton for +1,+2,+3 days).
Explain reliability diagram in the context of probabilistic forecast.
\item rewrite the definitions by first introducing TP, FP, TN, FN. Then explain the metrics both in words and with the formula
\item work on the baseline models paragraph. Explain that we are going to use persistence, a climatology model based on the number of flare-free days prior and the number of sunspots, a Naive bayes model and a logistic regression (based also on number of flare-free days prior and the number of sunspots). Explain that all of these are essentially zero-cost models.
Also add more narrative to explain the metrics in the context of a very imbalanced dataset. For instance, accuracy is not very relevant here, because a model that always predicts negative would have a very high precision (give the number)

\item flare free days and sunspots are not used separately but together
\item expand explanation of climatology, Naive Bayes and LR showing the math
\item The first subsection of Results will present the reliability curves for all 6 cases (M,X class) (24,48,76 hours in advance)
\item Note that there are only discrete values for forecast probabilities: M24 = 1,  5, 10, 15, 20, 25, 30, 35, 40, 45, 50, 55, 60, 65, 70, 75, 80,
       85, 90, 95 M48 = array([ 1,  5, 10, 15, 20, 25, 30, 35, 40, 45, 50, 55, 60, 65, 70, 75, 80,
       85, 90, 95]) M72 = array([ 1,  5, 10, 15, 20, 25, 30, 35, 40, 45, 50, 55, 60, 65, 70, 75, 80,
       85, 90, 95]) X24 = array([ 1,  2,  5, 10, 15, 20, 25, 30, 35, 40, 45, 50, 55, 60, 75]) X48 = array([ 1,  2,  5, 10, 15, 20, 25, 30, 35, 40, 45, 50, 55, 60, 70, 75]) X72= array([ 1,  2,  5, 10, 15, 20, 25, 30, 35, 40, 45, 50, 55, 60, 65, 75])
\item latex script to make 2x3 plots for reliability diagrams (each row one class, each column a different horizon)
\item write  a result section based on the following tables that summarize metrics for different models and both M-class and X-class and horizon of 24h,48h, 72h. This is done taking the standard threshold=0.5 for probabilistic forecast. [Tables in latex format omitted]
\item emphasize that SWPC is outperformed by any other model across the board, having sometime only the best accuracy and Brier which are both misleading for  imbalance datasets
\item write a section about the importance of "all clear" scenario. In order to test the ability of SWPC forecast to forecast all clear, we take the subset of days for which X-days-since-last-flare $>$ 30, that is days that have been clear for more than 30 days, in order to see whether the model has the ability to forecast a sudden X-flare. Mention importance for astronauts to be shielded.
In the subset there are 66 positives and 6992 negatives, thus this is an extremely difficult problem.
Using the optimal threshold for X-flares at 24 h (5\%) the confusion matrix is TN = 5735, TP = 56, FN = 10, FP = 1257. Draw conclusions
\item rewrite it. 15\% of the events were not correctly identified. Each one of this event could be lethal to an astronaut and so the model is unreliable and cannot be trusted
\item write a summary comment of the following tables. [Tables omitted] The thresholds for SWPC are very low, indicating a very badly calibrated prediction. Also, baseline models are consistenly better and FAR is always very high. Mention anything else that is useful for a full comprehension of those metrics and comparisons.
\item summarize the SWPC forecast methodology here:
The forecasts by NOAA/SWPC have long been considered a standard (Crown 2012) and have set the benchmarks against which methods are measured using the NOAA/SWPC event definitions (see commentary on this in Leka \& Barnes 2017). SWPC forecasters begin with a climatological approach. They classify the active regions and assign probabilities according to the historical flaring rates of different sunspot region classes (McIntosh 1990). (Note: SWPC's assignment of active region class is also considered "the standard.") From this, a forecaster may modify a region's probability according to region evolution, flaring trends, and forecaster experience and expertise. These region probability forecasts are combined for a full-disk forecast, which itself may be modified based on flaring history of recently rotated-off regions or indications of a highly active region about to return. Forecasters may also incorporate other model data when available. Initial forecasts issued at 22:00 (the Geophysical Activity Report and Forecast or RSGA) are valid beginning at 00:00 the next day. These are incorporated into the three-day forecast issued at 00:30, with a minimal but not zero probability of a forecast update in the intervening 2.5 hr.
\item make into a single paragraph sentence. focus on the methodology only
\item rephrase:
It is important to emphasize that the comparison against zero-cost baseline models is not intended to propose those baselines as alternative forecasting methods, but rather as the minimum an operational model needs to achieve to be worth using. In fact, one would hope that the official US forecasting agency would have produced, over the last 20+ years, a forecast that is not outperformed by statistical and zero-cost models.
\item Write the conclusion section. Highlight that users are currently using SWPC solar flare forecast assuming it has skills (being the official US govt forecast), when in fact it is no better than zero-cost baseline. In particular, the forecast is not useful in all clear scenario. We encourage the operazionalization of state of the art machine learning models that have proved skilled. Moreover, encourage SWPC to routinely run and publish validation of their models
\item write also an abstract and three one-sentence key points
\item Finally,write a plain-language summary.
This includes undefined/excessive acronyms, terms specific to your scientific field, obscure and unnecessarily long words, and words that have different meanings to non-scientists (e.g. modeling, positive feedback, mean, etc.). 
\item the paragraph about "all clear" has been more properly renamed "storm after the calm", while a "all clear" event is now an event when a X-storm happened on a day, and we want to forecast the likelihood that the next day will be 'all clear' (ie no X-storms)
\item the subset for the all clear scenario is as follows: +1, +2, +3 days after a X-class flare. For each day we focus only on the +24h forecast
\item Comment (TP = 150, FN = 1, TN = 38, FP = 576).
\item Add a Open Research Statement, which is a paragraph that explains where the data is available. The code for pre-processind and analysis (jupyter notebooks) is on a github repository

\end{itemize}

\noindent\textbf{The following prompt have been used for revising the manuscript (first round of review):}

\begin{itemize}
    \item write a section about the training strategy. In order to make the comparisons fair we do not use data that was not available at the time a forecast was issued to train statistical and ML models. We use the first 17 months as buffer (until 1998-01) and then re-train the models (climatology, logistic regression, naive bayes) monthly. In other words, every month uses a model trained with all the data available prior to that month. In principle this strategy is less relevant for climatology that, by definition, is supposed to be a statistical method that simply uses averages over a long time period, but for consistency we have used the same strategy for climatology as well. Persistence, on the other hand, does not require training
\item I also used a "baseline average" that is simply the average of climatology, naive bayes and logistic regression. Add a paragraph and explain the possible beneifts of averaging
\item I am also including persistence in the average
\item revise the paragraph that comments on M-Class flares based on the following three tables (24,48,72 hours ahead) [Tables omitted]
\item rephrase by emphasizing that accuracy and biar can both be misleading in imbalanced datasets
\item overall even persistence is either at par or only slighty worse than SWPC?
\item similar exercise writing a paragraph commenting the performance, now for X-class flares based on this table [Table omitted]. LR has been excluded because non-performing and BA is now the average of 3 models
\item Summary of forecast comparisons
\item I have now run the models using a different, optimized, threshold for binary classification. here below are the tables with metrics for M-class 24, 48, 72 hours ahead. [Tables omitted]
Write a susbection commenting those results.
\item including now results for X-class flares, write conclusions
\item what is the model that overall presents the best trade-off among all metrics?
\item summarize those findings in latex format
\end{itemize}

\end{document}